\documentclass[sigconf]{acmart}

\AtBeginDocument{%
  \providecommand\BibTeX{{%
    \normalfont B\kern-0.5em{\scshape i\kern-0.25em b}\kern-0.8em\TeX}}}

\copyrightyear{2021}
\acmYear{2021}
\setcopyright{rightsretained}
\acmConference[CHI '21]{CHI Conference on Human Factors in Computing Systems}{May 8--13, 2021}{Yokohama, Japan}
\acmBooktitle{CHI Conference on Human Factors in Computing Systems (CHI '21), May 8--13, 2021, Yokohama, Japan}\acmDOI{10.1145/3411764.3445522}
\acmISBN{978-1-4503-8096-6/21/05}

\usepackage{subfigure}

\begin{document}

% \title{Assessing Automated Decision Support for Domain Experts in a Clinical Annotation Context: Mediated Model Errors but a Loss of Agency}
\title[Assessing the Impact of Automated Suggestions on Decision Making]{Assessing the Impact of Automated Suggestions on Decision Making: Domain Experts Mediate Model Errors but Take Less Initiative}
\settopmatter{authorsperrow=4}
\author{Ariel Levy}
\authornote{Both authors contributed equally to this research.}
\email{aslevy@mit.edu}
\affiliation{%
  \institution{MIT CSAIL}
  \streetaddress{45 Carleton Street}
  \city{Cambridge}
%   \state{Massachusetts}
%   \postcode{02139}
  \country{United States}
}

\author{Monica Agrawal}
\email{magrawal@mit.edu}
\authornotemark[1]

\affiliation{%
  \institution{MIT CSAIL}
  \streetaddress{45 Carleton Street}
  \city{Cambridge}
%   \state{Massachusetts}
%   \postcode{02139}
  \country{United States}
}

\author{Arvind Satyanarayan}
\email{arvindsatya@mit.edu}
\affiliation{%
  \institution{MIT CSAIL}
  \streetaddress{32 Vassar Street}
  \city{Cambridge}
%   \state{Massachusetts}
%   \postcode{02139}
    \country{United States}
}

\author{David Sontag}
\email{dsontag@csail.mit.edu}
\affiliation{%
  \institution{MIT CSAIL}
  \streetaddress{45 Carleton Street}
  \city{Cambridge}
%   \state{Massachusetts}
%   \postcode{02139}
    \country{United States}
}

\renewcommand{\shortauthors}{Levy, Agrawal, et al.}

\begin{abstract}

Automated decision support can accelerate tedious tasks as users can focus their attention where it is needed most.
However, a key concern is whether users overly trust or cede agency to automation.
In this paper, we investigate the effects of introducing automation to annotating clinical texts\,---\,a multi-step, error-prone task of identifying clinical concepts (e.g., procedures) in medical notes, and mapping them to labels in a large ontology.
We consider two forms of decision aid: recommending which labels to map concepts to, and pre-populating annotation suggestions. 
Through laboratory studies, we find that 18 clinicians generally build intuition of when to rely on automation and when to exercise their own judgement.
However, when presented with fully pre-populated suggestions, these expert users exhibit less agency: accepting improper mentions, and taking less initiative in creating additional annotations. Our findings inform how systems and algorithms should be designed to mitigate the observed issues.
\end{abstract}




\maketitle

\section{Introduction}
Recent advances in machine learning have revolutionized many tasks by allowing human decision-makers to work in tandem with automated decision support.
Decision tools have emerged across a variety of disciplines including medicine \cite{wang2016deep}, sports \cite{Ono2019}, and criminal justice \cite{Kleinberg2018}. 
Such support is particularly valuable when the task is tedious or requires domain knowledge, since it is difficult to outsource and domain experts are often a scarce resource.  
The introduction of automation can decrease the cognitive load on human decision-makers, enabling them to focus their attention where it is most needed.
Since automated systems and humans often have complementary strengths, joint systems can outperform either alone \cite{wang2016deep}. 

However, while such hybrid intelligence systems are very promising, it is important to understand the dynamic between the human decision maker and the automated system, a topic of longstanding study \cite{Lee2004}. For optimal results, the human needs to understand when to listen to the computer and when to exercise their own agency \cite{Bansal2019}. If not, the human may develop a misplaced trust in the automation, which could have the adverse effect of degrading their output, as has been demonstrated in prior work \cite{Suresh2020}. Further, this trust may cause them to lose critical engagement with the task, so they do not have the attention to intervene when necessary, as has been shown in self-driving car examples \cite{Hergeth2016}. However, this phenomenon has been less studied for domain experts, who may be more confident in their own abilities and more skeptical of automation \cite{Liberati2017}. 

In this work, we use a clinical text annotation task to study whether for domain experts, their expertise mediates their interactions with automation. In particular, we  investigate whether domain experts display the misplaced trust and loss of agency that has been described in other past work, and see whether they remain critically engaged in the task at hand when using automation. Clinical text annotation is a complex, multi-step task and therefore a useful sandbox for these questions. As described further in Section \ref{sec:descr}, the task involves sifting through clinical notes to extract and map mentions of clinical concepts (e.g. symptoms, procedures), which are often written in overloaded jargon. For example, \texttt{MS} can refer to \textit{mitral stenosis}, \textit{multiple sclerosis}, \textit{medical student}, or \textit{mental status}. While challenging, accurate extraction of clinical concepts is crucial to enabling large scale retrospective studies over past electronic health record data. The first step in extraction involves identifying which spans of text refer to clinical concepts and need to be annotated. This requires users taking initiative to identify their own subproblems, allowing us to study their agency and critical engagement when automated aid is provided. The next step involves mapping each span to a concept label in a large (>400,000) medical vocabulary \cite{Luo2019}. Since the search space is so large, decision aid could decrease users' cognitive burden. 
However, recommendations could keep users from searching further when needed and decrease their accuracy, and this dynamic enables us to study the trust users build in automation.

We introduce a new text annotation system that provides two major forms of automated decision support: label recommendations for mapping spans to concepts, and fully pre-populated annotation suggestions. Our label recommendations surface a set of ten model-proposed labels after a user chooses a span of text to annotate. The user can choose to go with one of the recommendations or decide to search further. The pre-populated annotations suggest full sets of both text spans and corresponding labels. The clinical text annotation task has some terms that occur frequently and usually map to just a single label (e.g. `hypertension'); we present these terms as pre-populated annotations that users can easily accept, allowing them to focus their attention elsewhere. 

We run a two-stage user study on 18 clinicians from 9 different United States medical institutions, in which we artificially vary the extent and accuracy of both of the aforementioned  decision aids. In contrast to previous work with discrete, often synthetic tasks, our label space includes 400k+ concepts, allowing us to simulate when models are only slightly off (missing the correct label, but presenting similar adjacent labels), instead of scenarios where the model is entirely incorrect, reflective of real-world error modes. In doing so, we are able to pinpoint specific impacts of automation errors. We investigate how reliant users are on automation by examining how they deal with poor recommendations and the agency they show in annotating additional concepts. We analyze each annotator's behavior and outcomes over approximately eight hours of annotation each, allowing us to account for effects over long time scales like tedium. Through this analysis, our paper makes three contributions:
\begin{itemize}
    \item We find that domain experts (n=18) are sufficiently engaged to notice when system label recommendations are inadequate. Due to the size of the label space, we were able to measure how often they chose to search, an objective proxy for trust in the recommendations. Our users have a strong intuition of when to explore further, but if this intuition is violated and a correct label isn't present when they might expect it to be, they accept substandard label choices for spans they selected. 
    \item When presented with fully pre-populated annotations, we find that our domain experts are more hesitant to exercise agency. While they do change incorrect labels, they are slightly more hesitant to intervene with incorrect spans, and they demonstrate less initiative in the creation of additional spans. Moreover, through exit surveys, we find that they thought they were being thorough and do not note this shift in their own behavior. 
    \item We analyze error patterns of our domain experts and find that trust in automation correlates across suggested labels and spans. However, misplaced trust and loss of agency do not correlate with each clinician's prior demonstrated competency at the task. We also find that human error patterns differ from algorithmic error patterns, indicating the utility of combining the two.
\end{itemize}

Our results, compiled from over a hundred hours of logged interaction between clinicians and decision support, strongly indicate that domain experts can fall susceptible to risks in human-AI teams. Our detailed characterization of users' behavior can help inform the design of user-facing systems for data collection as well as the machine learning models that ingest that data.

\begin{figure*}[h]
  \centering
  \includegraphics[width=0.8\linewidth]{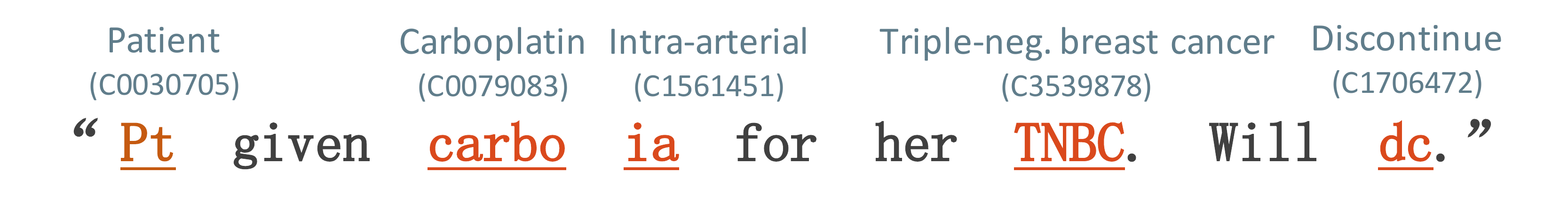}
  \caption{A typical example of a sentence found in a clinical note, displayed alongside the desired output of clinical annotation. The annotation task consists of both span identification (orange, underlined) and span mapping to concepts in the Unified Medical Language System (UMLS) vocabulary (teal, above). }
  \Description{An example sentence from a clinical note ("Pt given carbo ia for her TNBC. Will dc.") annotated with corresponding Unified Medical Language System (UMLS) clinical entities. The chosen spans are displayed underlined in orange, and the mapped concepts are shown in teal. Pt is mapped to patient, carbo to carboplatin, ia to intra-arterial, TNBC to triple-negative breast cancer, and dc to discontinue.}
 \label{fig:example_sentence}
\end{figure*}

\section{Related Work}
\subsection{Mental Models and Model Trust}
The idea that effective group work requires an accurate mental model of one's teammates---humans or AI---goes back decades \cite{Cannon-Bowers1993, Grosz1999}. In interviews, clinicians discussed their information desiderata when onboarding new AI decision support: strengths, weaknesses, point-of-views \cite{Cai2019}. These needs closely mirrored the information they use to create mental models of their colleagues when seeking second opinions. It has been empirically shown that users with better mental models of their AI teammates are more successful; in particular, knowledge of the error boundary enables the user to know when to trust automation and when to override it \cite{Bansal2019, Gero2020}.  However, in cases where users do not properly learn when to trust automation, they can become overly reliant on systems \cite{Lee2004}. On image recognition tasks, Suresh et al. found that people will trust an incorrect machine decision, even if they would have made the correct decision on their own \cite{Suresh2020}. Similarly, a study of radiologists showed that while decision aid helped lower-performing radiologists, it actually hindered the performance of the best ones \cite{Povyakalo2013}. Further work has shown that users' subjective evaluations of their trust in and relationship to decision support systems does not always align with actual outcomes, highlighting the importance of quantitative studies \cite{Bucinca2020}.   

Due to the importance of calibrating trust in decision aids, several recent studies have analyzed how the presentation of model decisions modulates trust.  They’ve investigated factors including reported model confidence, model explanations, overall model accuracy, and initial model accuracy \cite{Zhang2020, Yin2019, Papenmeier2019, Nourani2020}. In these studies, users choose to accept or reject the model’s output on discrete binary tasks and self-report their level of trust in the model. 

In this work, we study a more complex task which (i) has a large label space (>400k), requiring reliance on computational aid and allowing us to measure trust by how often users search, and (ii) requires users to select spans, additionally allowing us to measure agency. Further, the time scale of our study is longer than past work, allowing us to account for real-world effects such as tedium that would likely affect reliance and agency. The primary goal of clinical text annotation is to create data to better train machine learning models. However, if we provide an initial model as decision support and users become overly reliant on that model, their output could closely resemble the model's, instead of the underlying truth. This would lead to models being fed back their own outputs as inputs in future training; this could cause a dangerous feedback loop in which our models become even more confident in their own incorrect decisions \cite{Sculley2015}. Therefore, it is imperative to understand if our users will properly mediate model errors, or simply reflect them back.

\subsection{User Agency}
Some tasks like image classification have discrete inputs with clear objective outputs. However, for other tasks that require initiative or creativity, there is also the worry that humans will cede their agency to automated aid \cite{Heer2019}. For example, in the translation task, there are often multiple equally valid outputs. When translators used an interface that displayed machine-suggested recommendations, they noted that they ceded agency and would conform to the machine's recommended phrasing, even when it did not match their usual style \cite{Green2014}. Kulkarni et al. showed in an experiment on sketching, exposure to examples increases conformity of users' drawings \cite{Kulkarni2014}. Similarly, Siangliulue et al. investigated the effect of when examples were shown; users exposed to regularly-spaced examples produced the fewest ideas \cite{siang2015}.

In our work, clinical text annotation does not just require mapping mentions to concept labels, but it also requires deciding what terms are clinically relevant concepts and need to be labeled. While automated methods can provide users with a subset of these terms that need to be labeled, users could fall susceptible to similar patterns of tunnel vision, where their attention and mindset becomes fixed on what has been provided. On the other hand, a note contains over a hundred annotations, and a well-designed suggestion system could decrease the burden on users, enabling them to focus their attention where it is needed most, instead of replicating what is known.

\subsection{Annotation}
There are many ways to annotate, explore, and understand text corpora, but one common preprocessing step is to identify the entities within a document; in the general domain, these tend to be names, places, and organizations \cite{Brehmer2014}. For example, Jigsaw is a system for text exploration (e.g. articles, reviews); as a starting place, it algorithmically pre-tags entities in documents, and then allows users to make corrections to these annotations \cite{Gorg2013}. Historically, identifying clinical entities is particularly tricky, since the language in doctors' notes can essentially be considered its own dialect, and there are many overloaded terms, e.g. \texttt{MS}. Deciding what counts as a clinical entity (finding, disease, procedure, treatment, lab test) and what defines an entity's boundary is much more ambiguous than in the general domain; this is because concepts are not just proper nouns and often overlap with one another.

Due to the importance of understanding concepts and entities in text, there are a host of tools focused solely on entity annotation; however, these tools were created for tasks with a more limited number of labels (on the order of <30-40) that can all be simultaneously displayed.  Many of these tools (WordFreak, GATE, BRAT, WebAnno, Knowtator and YEDDA) do include system suggestions and pre-annotations \cite{Morton, Yang2018, Stenetorp2012, Kenter,yimam-etal-2014-automatic, ogren2006}. There have been multiple studies studying the impact of pre-annotation on outcomes, though the focus is often on efficiency gains \cite{yimam-etal-2014-automatic, Ganchev2007}. Past literature has also examined effects of pre-annotation on performance, both in standard NLP and clinical settings, but these have had mixed conclusions on the bias induced and the resulting time savings \cite{Lingren2014, Gobbel2014, Fort2010, B.R.2014}.  These works were based on small numbers of annotators (generally $n=2-4$) and focused primarily on overall agreement of an annotator with the gold standard and other annotators, rather than directly analyzing when they accepted incorrect aid. Further, our work differs due to (i) the large label space we are mapping to and suggesting over, which intensifies the dynamics around trust and mental models, (ii) the wider task definition (e.g. compared to part-of-speech tagging or identification of a few specific symptoms), and (iii) our purposeful introduction of certain modes of errors. 

There have been several open-sourced datasets consisting of our same clinical text annotation task, or slight variants on it \cite{Agrawal2020, Elhadad2015, Luo2019, Suominen2013, Pradhan2015}. These datasets have been manually created without decision support; annotators have used external websites to search the label space \cite{NIH-NLM2015}. As a result of the tediousness of the task and lack of specialized tooling, the datasets are small, with the largest being on the order of hundreds of notes.

\section{Clinical Annotation as a Case Study} \label{sec:descr}
Here, we describe the task of clinical text annotation in greater detail and describe why it serves as an interesting testbed for investigating trust and agency in the presence of automated decision aid.

Clinical notes are an incredibly rich source of data about a patient's interactions with the healthcare system. They are lengthy, containing detailed information about a patient's state, their underlying conditions, any procedures performed, and a wealth of other data that may be absent from the structured fields of the patient's electronic health records. This data has the potential to power large scale clinical endeavors; use cases include automatically matching patients to clinical trials, learning patterns of adverse drug events in the real-world, creating cohorts of patients for retrospective outcome studies, and discovering how disease presentation differs among subpopulations. Unfortunately, idiosyncrasies of clinical text can make notes difficult even for clinicians to read, since there is overloaded terminology and frequent use of shorthand, which often differs across medical specialties and institutions. Clinical text can be incomprehensible to laypeople, making it difficult for them to understand their doctors' notes about them, and to natural language processing (NLP) models, which are primarily trained on datasets built on standard text.

As an example of why clinical text can be convoluted, let us consider the sentence \texttt{"Pt given carbo ia for her TNBC. Will dc"}, which is shown annotated in Figure \ref{fig:example_sentence}. Clinical concept annotation involves two steps: (i) identifying the spans in text that correspond to clinical concepts (underlined in orange) and then (ii) mapping those spans to structured medical vocabularies (shown in teal). The latter involves understanding that \texttt{Pt} refers to a \textit{patient} not a \textit{physical therapist}; that \texttt{carbo} is shorthand for \textit{carboplatin} and not \textit{carbo-dome}; that \texttt{ia} refers to an \textit{intra-arterial} route of injection not an \textit{intra-articular} one; and \texttt{dc} refers to \textit{discontinuation} of a drug, not \textit{discharge} from the hospital or a \textit{Doctor of Chiropractic}. Additionally, spans can overlap; consider the phrase \texttt{dirty UA}. As a whole, it refers to the finding \textit{urine screening abnormal}, but \textit{UA} alone refers to the procedure \textit{urinalysis}. Deciding what is purely a descriptor and what is part of a concept can be difficult in edge cases. Further complicating the process, there are often only subtle nuances between different clinical concepts in vocabularies, so the mapping from spans to concepts is not always one-to-one. There are occasionally multiple equally correct concepts (e.g., \textit{sputum} and \textit{sputum culture}), no correct concepts, or only approximately correct concepts. Therefore, extracting the valuable information trapped in free text clinical notes requires clinical domain expertise to disambiguate from context, but this is prohibitive at scale and limits opportunities for research. 

As a result, the clinical NLP community endeavors to build systems that can automatically identify and disambiguate mentioned clinical concepts---e.g. conditions, symptoms, medications, and procedures---for use in downstream research and future care. 
Entities are typically mapped to SNOMED Clinical Terms, a subset of the Unified Medical Language System (UMLS) that contains over 400,000 concepts, or to RxNorm, a comprehensive clinical drug vocabulary ~\citep{NIH-NLM2015, Nelson2011}. 
Each concept in the vocabulary is accompanied by an (incomplete) set of synonyms and a set of categories to which it belongs, including \textit{Signs and Symptoms}, \textit{Therapeutic Procedure}, and \textit{Medical Device}. 
While there are existing computational systems to conduct this clinical concept extraction, they are not yet robust enough for reliable clinical use, extracting only about half of the concepts in a note fully correctly \cite{Agrawal2020}. 
This half largely consists of common concepts that algorithms have seen many times (e.g. diabetes, hypertension). 
However, existing algorithms often fail to extract concepts that can be described in a multitude of ways, like procedures, or rarer concepts, despite these being critical for many of the downstream applications. 
This middling algorithmic performance can be partially attributed to the dearth of annotated clinical text that can serve as training data, due to the tediousness of the task and the lack of specialized tooling.

\begin{figure*}[]
  \centering
  \includegraphics[width=\textwidth]{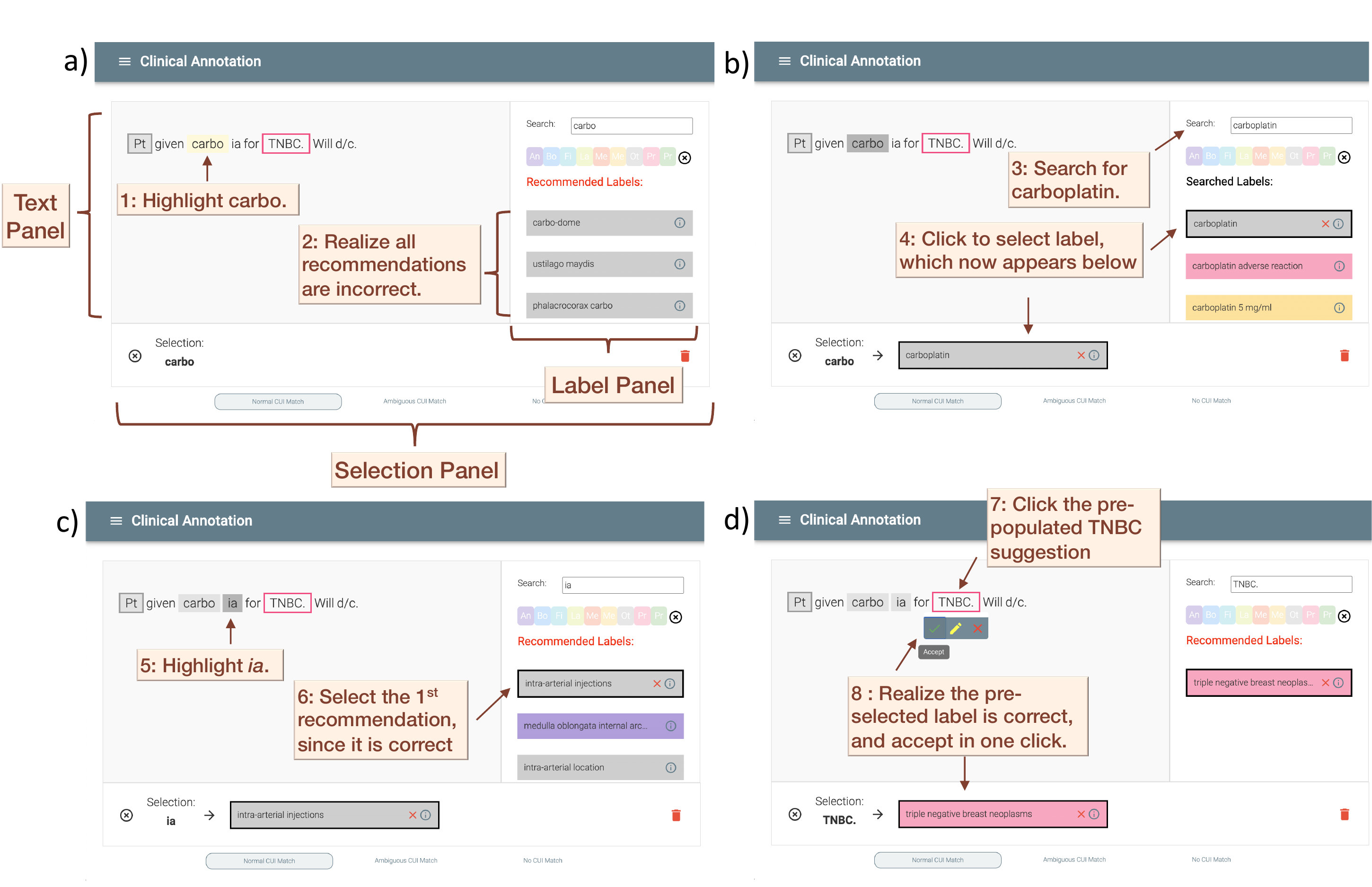}
  \caption{Our clinical annotation system in the midst of annotating our sample sentence. Panel a) shows the text, label, and selection panels. In the above, \textit{Pt} has already been annotated; the user annotates \textit{carbo} using the search feature in panels a) and b), \textit{ia} using the automatically surfaced label recommendations in panel c), and \textit{TNBC} using a pre-populated annotation suggestion in panel d).}
  \Description{The figure consists of 4 subfigures which are screenshots of our annotation system, and these correspond closely to the description in the walkthrough in Section 4.1. The running sample sentence is shown in the middle in the text panel, the label panel takes up the right half, and the selection panel is on the bottom. In panel a), the word carbo has been highlighted, and on the right in the label panel, a series of incorrect options are displayed (e.g. carbo-dome). In panel b), the user has typed carboplatin into the search bar in the label panel, and the correct label is displayed at the top of a list. This label is shown selected, and it also now appears in the selection panel. In panel c), the user has highlighted ia, and the recommended labels in the label panel have the correct answer displayed first. The user selects it, and again, it automatically then appears in the selection panel. Finally in panel d), we notice that TNBC has had a pink outlined box around it the whole time. This box is clicked on, and three small icons appear underneath: a green check for acceptance, a yellow pencil for modification, and a red X. The selection panel comes pre-populated with the correct label before the user does anything, and the panel indicates that clicking on the green check will complete the annotation.}
  \label{fig:system}
\end{figure*}

To address this gap, we introduce a platform built specifically for concept annotation; the task is ripe for decision aid since it is tedious, can only be conducted by domain experts, and existing systems can act as partial solutions. The task acts as a good sandbox since users' decisions of what to label enable us to study how decision support affects their autonomy and agency, and the large label space enables us to study how users build trust in recommendations.

\section{System}

Due to the complexities and idiosyncrasies of the clinical annotation task, we built a custom  annotation platform \footnote{An up-to-date version of our platform can be found at clinicalml.github.io/prancer.}. The platform is based  on top of React, Typescript, Node.js, and Flask. In this section, we first present a walkthrough of representative usage of the platform, and then elaborate on the system behind automated decision support.
\subsection{Usage Walkthrough} \label{walkthrough}
The platform consists of three main panels, shown in the top left of Figure \ref{fig:system}: a text panel, a label panel, and a selection panel. Users begin by looking at the text displayed in the text panel; in  Figure \ref{fig:system}a), we see that the shaded box around \texttt{Pt} indicates it has already been annotated, and the user highlights the next clinical span, \texttt{carbo}. Upon highlighting, the selection panel shows that \texttt{carbo} has been chosen, and the label panel shows a list of automatically generated recommendations. The user notices that none of the provided options are correct, then in Figure \ref{fig:system}b), uses the search bar to explicitly search for \textit{carboplatin} instead. A list of search results appear, color-coded by concept type (e.g., pink for \textit{Problems}, grey for \textit{Other}); we use the 11 concept types described in \cite{Patel2018}. Our user would notice that the first suggestion is correct and click on it, placing the label in the selection panel. The user could continue searching for additional labels, or move on. In Figure \ref{fig:system}c), they highlight \textit{ia} and the first displayed label is correct, so they can select it and move forward. In Figure \ref{fig:system}d), the user then comes to \texttt{TNBC}, which already has a pink outlined box around it, indicating that the system has pre-populated an annotation for \texttt{TNBC}. Upon clicking the box, the system displays what it has identified as the likely correct label, which it auto-populates in the selection panel. The user then can choose to accept this span and label combination in a single click on the green checkbox that appears as a dropdown; alternately, the user could choose to modify the label via the yellow pencil or to discard the result via the red `X'. At any time, the user can return to edit any annotation by clicking on its surrounding box.

 We adopted the multi-panel view to mimic other interactive writing/tagging applications (e.g., Grammarly) where text spans are highlighted in situ to indicate the presence of recommendations, which are then shown in a side panel \cite{Inc.Grammarly2018}. In earlier prototypes, we explored alternate design decisions. For example, we considered displaying recommendations in-line via a drop-down menu and displaying existing labels in-line; we found that the former idea obscured surrounding text that was useful for ascertaining context, and  the latter cluttered the screen with minimal benefit. We also explored alternate recommendation confidence indicators and suggestion utilities such as \textit{Recently Used}, and our final interface was a synthesis of these ideas. 
 
\subsection{Additional Feature Details}
In addition to the features explicitly walked through above, there are a few more to accommodate the challenges of clinical text annotation. Users can indicate in the selection panel if there is no matching concept, or only an ambiguous match. Further, users do have the ability to select overlapping spans, and when there are multiple annotations for a single word, users may toggle between these annotations. If a user needs more descriptive information on a concept, they can click the \textit{(i)} button on any label in the label panel to surface the official medical definition. Finally, a user can click on one of the colored boxes under the search bar to filter recommendations or search results to one of the 11 concept types; these colors correspond to those used in the search and in the annotation bounding boxes.  

In the platform, any addition, deletion, or modification of manual or suggested annotations is automatically saved in a JSON-serialized dataset. Saved features include the character span numbers, the labels selected, the timestamp, and whether the annotation was manual or suggested. For the purposes of our study, we also log all interactions the user has with the platform. 

\subsection{Automated Decision Aid}
As introduced in the walkthrough, we introduce automated decision aid into our platform in two major ways. The first is via automated label recommendations given a user-highlighted span (see \textit{ia} in Figure \ref{fig:system}c); the second is via fully pre-populated annotation suggestions (see \textit{TNBC} in Figure \ref{fig:system}d). We describe each modality and its motivation in greater depth below.

\begin{figure*}[h]
  \centering
  \includegraphics[width=0.7\linewidth]{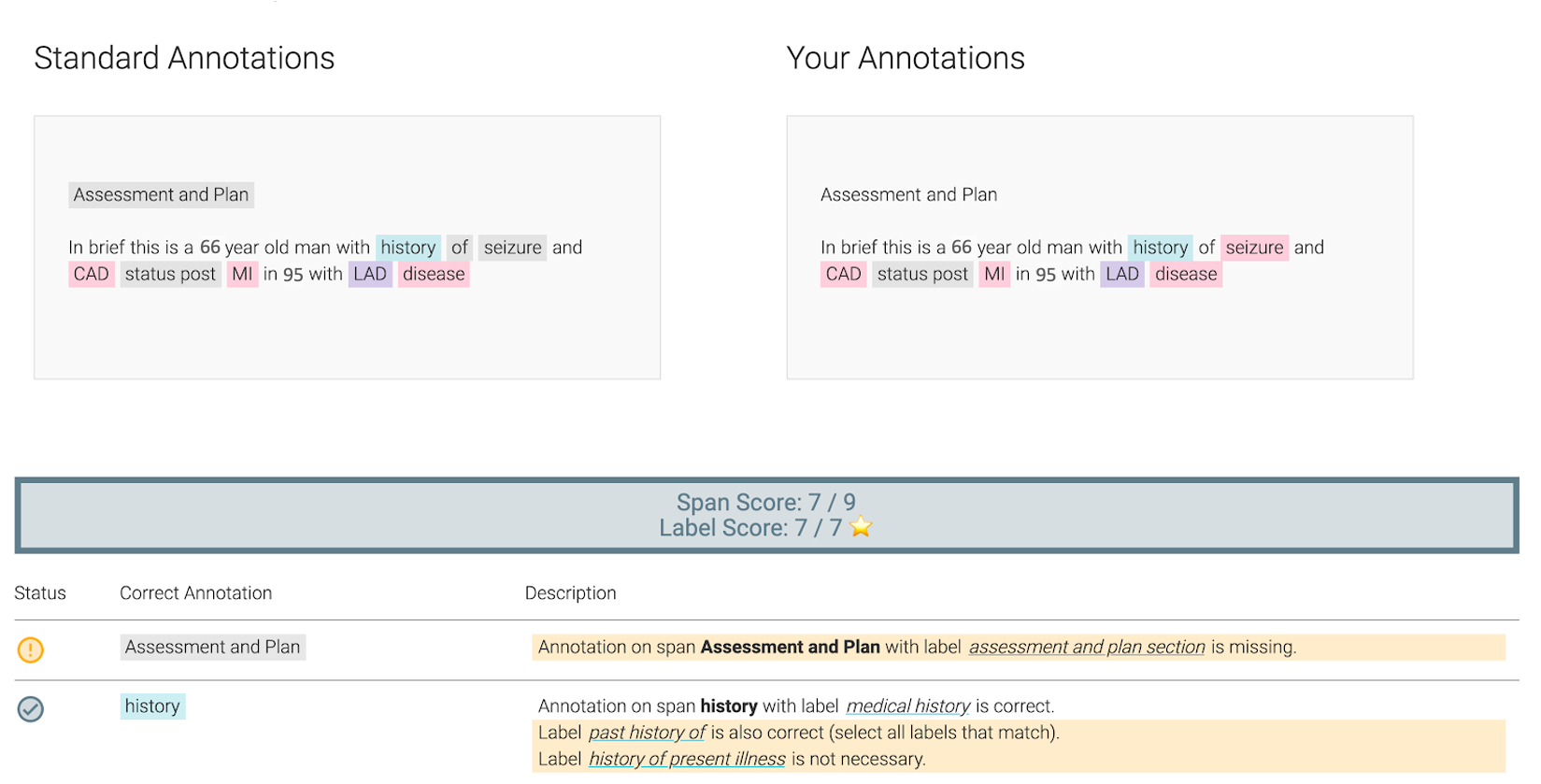}
  \caption{Tutorial mode built into the platform to train users with rounds of iterative feedback. Gold standard and chosen annotations are displayed side-by-side, with feedback on all annotations provided below.}
  \Description{This is a screenshot of the tutorial mode of the platform. At the top left, the user is presented with an image of what a perfect labeling would have been, and on the right, they see their own annotations. Directly below, they receive a score for their span recall (7/9), and for their label accuracy (7/7). Below, the discrepancies for each annotation are listed, if any. Two are displayed here. The top one has a yellow exclamation mark and states "The span Assessment and Plan with label Assessment and Plan Section is missing." The second has a grey check mark and alerts the user that they had a correct answer, but one of their extra labels may be erroneous; this does not detract from the user score.}
  \label{fig:tutorial}
\end{figure*}

Automatic label recommendation could greatly decrease the amount of searching users have to do. While there is a large label space (>400,000 concepts), as discussed in Section \ref{sec:descr}, there is also a set of lower-hanging fruit. These are text spans that, once highlighted, are either straightforward to label or narrow down to a set of labels.  On their clinical text corpus, Luo et al. showed that when provided with a correct text span, simple heuristic methods can achieve 77\% label accuracy; the highest accuracy yet achieved on their data set is 85\% \cite{Luo2019}. Therefore, while existing algorithms are not sufficiently robust to conduct automatic extraction in clinical workflows, given an identified span, they are sufficiently advanced to narrow down to a set of useful recommendations in the vast majority of cases, particularly for frequently occurring concepts. As a result, our system surfaces recommendations automatically on the right label panel when a user highlights a span, as seen in Figure \ref{fig:system}(a, c). On the backend, the recommendation system is built within a Python wrapper, allowing for easy extensibility to any modern machine learning model as algorithms improve. In our studies, we use the recommendation algorithm detailed in Agrawal et al., which draws on classical information retrieval techniques from NLP \cite{Agrawal2020}. 

Our platform can also incorporate fully pre-populated annotations, where both the span and the label are provided. As described in Section \ref{sec:descr}, the field of clinical NLP has already created several systems that attempt to automatically extract concepts end-to-end: both identification of spans and mapping those spans to labels.
While these are obviously imperfect, thus necessitating the need for clinical text annotation, they can be used to recover at least half of concepts. 
Given that a note can easily have hundreds of clinical concepts, these systems could relieve the burden on the human user by potentially allowing them to focus their attention on more difficult cases. In our platform, pre-populated annotations appear as outlined boxes around the suggested span, as seen with \texttt{TNBC} in Figure \ref{fig:system}. On the backend, the pre-populated annotations are computed and rendered before the user begins their annotations. Therefore, latency is not a concern, allowing any concept extraction system to be used to provide the pre-populated annotations. Possible existing concept extraction systems include cTAKES, MetaMap, MedLEE, and scispaCy \cite{Aronson2010, Savova2010, friedman1995medlee, Neumann2019}.

Finally, due to the repetition of terms within a note, we additionally implemented a feature that propagated concept annotations to repeat occurrences of the same entity. For example, if a user marks \texttt{carbo} at the beginning of the note as \textit{carboplatin}, all future occurrences of \texttt{carbo} in that note will appear with a pre-populated annotation for \textit{carboplatin}, that users can again choose to accept, modify, or delete.

\subsection{Tutorial}

Finally, we included a tutorial mode in our system, shown in Figure \ref{fig:tutorial}. As input, the tutorial mode takes in a series of snippets and their gold standard annotations. To begin, users annotate the first snippet using the standard interface shown in Figure \ref{fig:system}. After, they are brought to the tutorial screen shown in Figure \ref{fig:tutorial}. It presents the gold standard annotations on the left, the user's annotations on the right, and below, it provides a score based on the number of spans correctly recovered by the user, as well as the number of correct labels. Underneath, it iterates through all the annotations, providing a description of the differences between the gold standard and the user output, if any. Then, the user iterates through steps of annotation and tutorial feedback until the mode is over. The tutorial serves two purposes: (i) to familiarize users with the annotation task and the platform and (ii) to allow them to form mental models of the accuracy of the automated features.

\section{Stage 1: Label Recommendations}\label{sec:stage_1}
In this first stage of the user study, we test the effect of presenting automatic label recommendations once a user highlights a span of text. Recommendations could decrease the burden of the user searching over the large label space, especially since algorithms could present a correct label over 80\% of the time \cite{Luo2019}. However, there is a concern that in the presence of recommendations, users will become overly trusting and accept substandard labels instead of taking the initiative to search further. This could lead to a feedback loop if the data created were used for updating the automation model. Therefore, in this stage, we investigate whether our domain experts form appropriate intuition of when to search further, or whether they become complacent in the presence of recommendations. 
\subsection{Experimental Design}
To ensure a consistent and sufficient clinical background, we required users to have completed at least two years of medical school and to have experience with clinical notes in United States healthcare settings. We recruited 18 clinicians via Twitter and email lists. Our users (8 men, 10 women) consisted of 4 medical school graduates, 6 fourth years, and 8 third years; they came from 9 different medical institutions across the United States.  Users were compensated \$20 per hour for their time, and each spent between 4 and 5 hours total on this stage. All components of the study were conducted virtually and were ruled IRB exempt.

\begin{figure*}%
\centering
\subfigure[Total recall compared to the gold standard. Recall shown over all annotations, ``Difficult" annotations (where recommendations do not surface a correct answer), and "Weakened" annotations (where the correct answer was removed from \textit{Weakened} mode).]{%
\label{fig:recall}%
\includegraphics[width=3in]{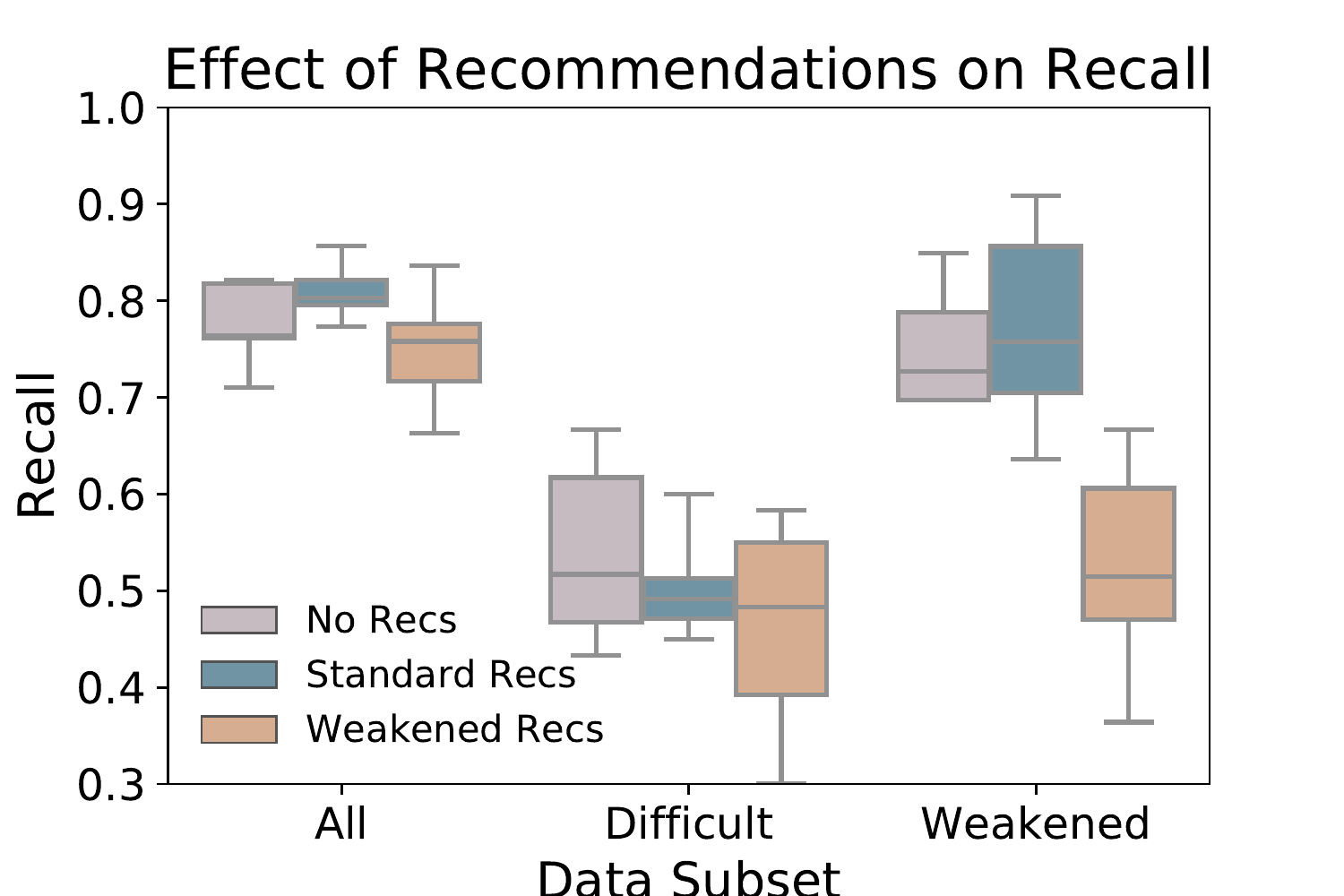}
}%
\qquad
\subfigure[Histogram and density curve for time taken to label a span under different recommendation modes (median of 6 seconds in \textit{None} and 3 seconds in \textit{Standard} and \textit{Weakened}).]{%
\label{fig:timing}%
\includegraphics[width=3in]{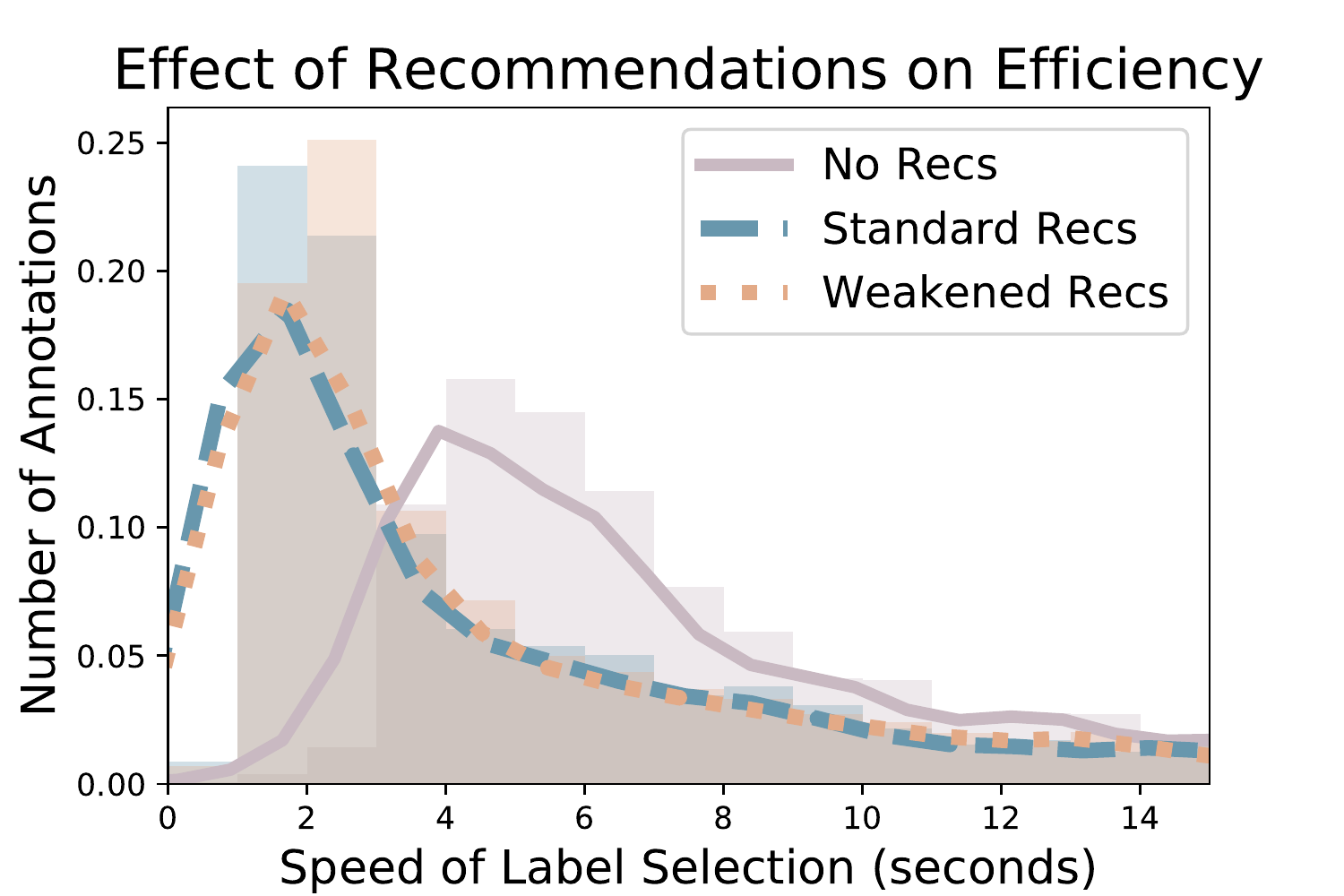}
}%
\caption{Accuracy (total recall) and efficiency (time to label) results for users with label recommendations (\textit{Standard} and \textit{Weakened} modes) and users without (\textit{None} mode).}
\Description{This figure consists of two subfigures. On the left, there are box-and-whisker plots corresponding to total recall on All annotations, Difficult annotations, and Weakened annotations. These measures are given individually for all three modes. On all annotations, recall for all three modes is similar, centered around 0.75 to 0.8 recall. On Difficult annotations, there is a wider range of performance between annotators (from 0.3 to 0.65), but the medians across all three modes are about 0.5. Finally, for the Weakened annotations, which only affect the weakened mode, users in that mode do significantly worse (median of 0.5 compared to about 0.75). On the right, we see a histogram overlaid with density curves, showing the estimated speed of label selection with recommendations and without. For both Standard and Weakened Modes, the median is three seconds and peaks around 2 seconds. For the No recommendation mode, the median is 6 seconds and peaks around 5.}
\end{figure*}

Our study had a multiple factor design, where we evaluated the performance of users (i) across different recommendation modes and (ii) across different annotation task difficulties that naturally arose in the task. First, users were assigned randomly to one of three modes: the \textit{None} mode, with no recommendations (5 users), the \textit{Standard} recommendation mode (6 users), or the \textit{Weakened} recommendation mode (7 users); the number of users per mode differed due to unanticipated changes in clinicians' schedules. To isolate the effect of recommendations, we did not include any pre-populated annotations in this stage. The \textit{Standard} mode presented the model recommendations in all cases. The \textit{Weakened} mode presented the same recommendations as the \textit{Standard} mode, but with the correct label removed in 25\% of the nontrivial examples. Examples were considered nontrivial if the text did not directly match a concept or any of its synonyms in the medical vocabulary. In other words, we would only remove the correct recommendation if there would be a better search query to find the concept; the removal of a concept was done consistently for all instances of a term across this stage. In all, a correct label was presented in one of the ten displayed recommendations 83\% of the time in the \textit{Standard} mode and 73\% in the Weakened mode.
 Across all modes, we used the search provided by the Unified Medical Language System API \cite{Harbourt1993}. 

To train users, we individually gave each of them a 30-minute presentation, detailing the annotation rules they should follow, taken from \cite{Agrawal2020}. Users were encouraged to interject and ask for clarifications, as needed. Then, we had them annotate a sequence of eight clinical snippets using the tutorial in Figure \ref{fig:tutorial}, while we were present. Each user conducted the tutorial in their assigned study mode, so that they could be introduced to the strengths and weaknesses of the automation in their mode. We ensured the tutorial modes had approximately the same proportions of correct and incorrect recommendations as the full notes to allow users to build appropriate mental models.

Users then tagged 3 sections from de-identified clinical notes from the MIMIC-III Critical Care Database \cite{Johnson2016}. The order of notes was randomized between users. Each user annotated the first note of each stage live over Zoom, talking out loud, and then annotated their next two notes asynchronously, with screen recording. Following each stage, users received a followup survey asking about their workflow, specifically their confidence in and reliance on the automated decision aid features. 

For evaluation, we compared to the gold standard released by Agrawal et al., which contained 335 annotations over the sections used in this stage \cite{Agrawal2020}. Since users are allowed to select multiple labels, we treated an annotation as correct if any selected label were among the gold-standard labels. Further, since there may be multiple correct labels and the gold standard may not have contained them all, we manually checked whether any additional user labels were correct, and if so, added them to our gold standard. This process was conducted blind to the user and their mode. Further, we excluded all spans where the label was considered ambiguous in the gold standard. 

We analyze several dependent variables: users' accuracy, the speed of clinical annotation, and the actions taken in the platform (e.g. choosing to search). Accuracy was evaluated via \textit{span recall}, the proportion of annotations in the gold standard that users annotated the span for, \textit{total recall}, the proportion of annotations in the gold standard that users got the correct span \textit{and} correct label for, and \textit{label accuracy}, the percentage of time users chose a correct label for a set of spans. In our mixed effects design, our other factor was annotation difficulty.  In evaluating user accuracy, we consider multiple subsets of annotations, including ``Easy'' examples (the examples in which the correct label is provided in both recommendation modes), ``Difficult'' examples (the examples in which the correct label is not provided in the \textit{Standard} mode), and ``Weakened'' examples (the examples in which a correct label is provided in the \textit{Standard} mode but not in \textit{Weakened}.) We also examine performance across clinical concept types.

\subsection{Results}

For our multi-factor design (user mode and annotation difficulty), we first conduct an Aligned Rank Transform (ART) procedure \cite{Wobbrock2011}. We find that total recall is significantly affected by mode (p<0.03), annotation difficulty (p<1e-16), and their interactions (p<1e-4). Both factors and their interaction additionally significantly affect how often users search on the platform (p<1e-8 across all three). We now deep-dive into pairwise comparisons and implications behind results. 

\subsubsection*{Recommendations increase annotation efficiency and seem to decrease tagging fatigue. \\}
Across all annotations, users in both recommendation modes are able to find labels far quicker (median of 3 seconds) than users without recommendations (median of 6 seconds) as displayed in Figure \ref{fig:timing}; the median time for both recommendation modes is statistically significant faster than users without (p<0.05 across both, adjusted two-sided Mann-Whitney U test).  When a correct answer is provided in the ten displayed recommendations, users only require a median of 2 seconds. On the set of "difficult" examples where users are provided only with incorrect recommendations, they take a median of 10 seconds, the same as those without recommendations take on the "difficult" set. Further, the users in \textit{Standard} mode create an average of 12\%  more annotations than those in \textit{None} mode (375 vs 337), a statistically significant increase (p<0.02, two-sided Mann-Whitney U test). Users in \textit{Weakened} mode were in the middle of both, with an average of 353 (not significantly different). We hypothesize that the decreased workload stemming from recommendations led to a lower cognitive load for users, decreasing their tagging fatigue and enabling them to create more annotations.

\subsubsection*{Recommendations generally improve recall, and domain experts step in appropriately in spaces where algorithms fail. \\}
As seen in Figure \ref{fig:recall}, users in the \textit{Standard} mode had higher total recall over users in the \textit{None} and \textit{Weakened} modes (80\% vs. 76\% and 76\% at median, respectively). The superiority of the \textit{Standard} mode over the \textit{None} mode is not statistically significant, but results indicate that the presence of recommendations does not decrease recall, a prior worry. On the set of difficult examples (where the recommendations do not contain a correct label), users across all modes have approximately the same total recall (a median of 52\% for \textit{None}, 49\% for \textit{Standard}, and 48\% for \textit{Weakened}). One user noted in their survey that they “really appreciated the suggested labels, but ... these can institute bias due to availability”, the only user to mention such concerns. This user also had the highest label accuracy on the set of ``Difficult" terms which may hint that active awareness of bias can help combat it.

We observe that human recall is much more consistent across different types of clinical concepts; this stands in stark contrast to algorithmic methods. For example, algorithmic accuracy of one existing extraction system is around 48\% overall, but only 24\% for procedures \cite{Agrawal2020}. Meanwhile, humans achieve about 70\% accuracy on procedures, compared to approximately 80\% overall. This indicates that human errors don't follow the same error patterns as algorithms, and therefore are adding valuable signal.

\subsubsection*{Users develop intuition of when the recommendations should surface a correct answer, but label accuracy suffers when that intuition is disrupted. \\}

Here, we investigate whether users recognize when a correct answer is present in the provided recommendations, and when they need to search further. We break down percentages by mode and example type in Table \ref{tab:freq}. On ``easy" examples (those where a correct answer was in the recommendations), users in both the \textit{Standard} mode and the \textit{Weakened} mode only searched further 17\% of the time and ultimately chose a recommended label 96\% of the time. However, on ``difficult" examples, \textit{Standard} users searched further 85\% of the time and \textit{Weakened} users 82\% of the time. Therefore, we see that their search patterns are closely aligned, and they generally learn to search further when necessary. This indicates that users had a strong sense of cases in which the algorithm isn't surfacing the correct answer.  

However, we observe that performance breaks down if there is a "violation of intuition", namely users expect a correct label to appear and it isn't present due to synthetic removal; again we only removed answers for a random subset of ``nontrivial" examples, where there was no direct match, and a better term could be found by searching. As evidence, we examine the performance of users in the \textit{Weakened} mode on the ``weakened" examples, nontrivial examples in which the correct labels were randomly excluded from the recommendations. On these ``weakened" examples, a set of examples in which they might have expected the recommendation algorithm to surface the correct answer, they only conducted a further search 73\% of the time, and as a result, achieved significantly worse total recall (average of 53\%) than the other two modes (76\%), as seen in Figure \ref{fig:recall}. They searched significantly less here than on the "difficult" examples (p<0.05, two-sided Mann-Whitney U test). This difference indicates that users were not solely searching further based on whether they thought a label was missing, but also based on whether they thought the recommendation algorithm should have been able to surface the correct label.

\begin{table}[]
\begin{tabular}{|l|l|l|l|}
\hline
\textbf{\begin{tabular}[c]{@{}l@{}}Mode / \\ Data Subset\end{tabular}} & \textit{\begin{tabular}[c]{@{}l@{}}``Easy"\\ Examples\end{tabular}} & \textit{\begin{tabular}[c]{@{}l@{}}``Difficult"\\ Examples\end{tabular}} & \textit{\begin{tabular}[c]{@{}l@{}}``Weakened"\\ Examples\end{tabular}} \\ \hline
\textit{Standard}                                                      & 17\%                                                                & 85\%                                                                     & 15\%                                                                    \\ \hline
\textit{Weakened}                                                      & 17\%                                                                & 82\%                                                                     & 73\%                                                                    \\ \hline
\end{tabular}
\caption{The percentage of times users chose to initiate a search. Across modes, searches are rarely initiated when the true label is provided in the recommendations (``Easy"), and are often initiated when the true label isn't provided (``Difficult"). When a user expects a label to be provided but it is not (\textit{Weakened} mode on ``Weakened" examples), users search less (73\%) than they did when they didn't expect the label to show up (82\%).}
\Description{A table of the percentage of times users with recommendations chose to initiate a search across modes and subset of examples. For Easy examples, users in both modes search 17\% of the time. For Difficult examples, users in the Standard mode search 85\% of the time, and users in the Weakened mode search 82\% of the time. Finally, for Weakened examples, users in the Standard mode search 15\% of the time, and users in the Weakened mode search 73\% of the time.}
\label{tab:freq}
\end{table}

Therefore, on the examples where the \textit{Weakened} users did not find a correct label in the recommendations, but expected to find one, they were less motivated to search. In these cases, they tended to accept substandard labels they perhaps wouldn't have otherwise. Common failure modes include assuming there must be no matching label and (i) choosing that there is no label present (e.g. a user for \texttt{Diastolic CHF}), or (ii) settling for a related but suboptimal label (e.g. \textit{ultrasonography} for \texttt{echo} instead of searching for \textit{echocardiogram}). A \textit{Weakened} user chose that there was no code for 4\% of \textit{Weakened} examples at median, and 6\% on average. 

This decreased search behavior leads to significantly worse results. A pairwise Mann-Whitney U test shows that while total recall was not significantly different between modes for ``Easy" or ``Difficult" examples, the ``Weakened" mode was significantly worse at the ``Weakened" examples than the other two modes (p<0.02 for both); see Figure \ref{fig:recall}. 

\section{Stage 2: Annotation Suggestions}\label{sec:stage_2}
In this second stage of the user study, we study the effect of presenting a set of automatically pre-populated annotation suggestions, as described in Figure \ref{fig:system}, that users can choose to accept, modify, or reject. Since there are existing systems that could pre-annotate at least half of the data, it has been suggested that starting from scratch might be an unnecessarily tedious exercise. However, there are fears users might lose engagement in the task and as a result, accept annotations with incorrect spans or incorrect labels, when models are imperfect. As a potential mitigation, we explore whether informing users that provided labels are high or low-confidence will make them more attentive when most necessary. An additional potential worry is a loss of agency; unlike in vanilla classification tasks, the user has to select which regions of the text to annotate. If the text comes partially pre-annotated, users might be less likely to take the initiative to annotate further. Both of these concerns could cause a feedback loop in which future models trained on this data become even more confident in error modes.

\begin{figure*}[]
  \centering
  \includegraphics[width=0.7\linewidth]{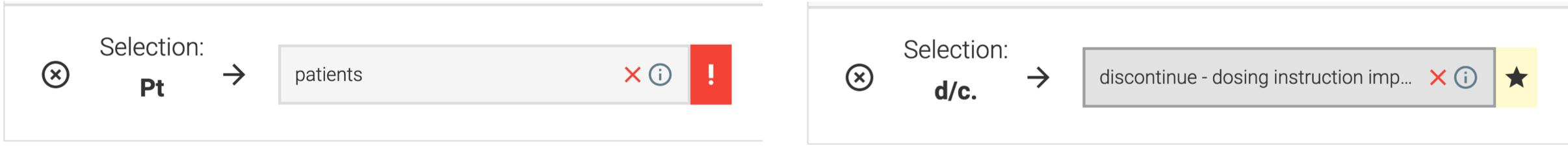}
  \caption{Flag indicators for suggestion confidence, displayed in the selection panel. Low-confidence (left) is indicated by an exclamation point on a red background, and high-confidence (right) is indicated by a star.}
  \Description{A pre-selected label presented with a tag with a star on a yellow background, indicating high-confidence, and another pre-selected label presented with an exclamation on a red background, indicating low-confidence from the model.}
  \label{fig:confidence}
\end{figure*}

\subsection{Experimental Design}
The same 18 clinicians participated in this stage of the study, and they spent approximately three hours each on this stage of the study. They were now re-assigned randomly to one of four modes: the \textit{No} suggestion mode (3 users), the \textit{Standard} suggestion mode (4 users), the \textit{Augmented} suggested mode (5 users), and the \textit{Weakened} suggestion mode (5 users). All modes were provided with the automatic label recommendations from the \textit{Standard}  mode in Stage 1; the back-end search was also updated to that same algorithm, since we are no longer directly testing recommendation versus search. 

The \textit{Standard} suggestion mode displays suggestions for the examples where there was an exact match between the example text and one of the synonyms in the medical vocabulary (76\% of the annotations in the gold standard). Of these suggestions, all spans and 85\% of the labels are correct. The \textit{Augmented} mode shows the exact same suggestions but includes a small flag on the label (see Figure~\ref{fig:confidence}), indicating whether the algorithm has high-confidence or low-confidence in the label. Approximately one-third of the suggested labels are considered low confidence; the high-confidence labels are 90\% correct, and the low-confidence labels are 70\% correct. The \textit{Weakened} suggestion mode has the same setup as the \textit{Standard} suggestion mode, but it also includes an additional set of 21 suggestions over incorrect spans. To mimic real-life algorithmic errors, these spans were taken from real incorrect span outputs of the clinical extraction systems cTAKES and MetaMap~\cite{Aronson2010, Savova2010}.

Since users already had some exposure to the tool and annotation schema, they completed just four steps in the tutorial this stage, again receiving feedback iteratively after each step. The tutorial sentences were designed to contain span and label errors in the same proportion as the full study, so that users could understand the role of automated decision aid. As in the previous stage, they again labeled three notes each, which contained a total of 449 spans in the gold standard.

The evaluation is conducted as in the first stage, using the same metrics and comparing to the same gold standard. In this case, we also analyze the subsets of annotations across suggestion confidences and provided label and span accuracies. Where appropriate, we also compare our users' outcomes in Stage 2 to their corresponding outcomes in Stage 1, to understand individual shifts in behavior. 

\subsection{Results}

\subsubsection*{Users were relatively accurate at assessing the correctness of the labels for pre-populated suggestions. Accuracy differed widely across users, but was not a function of skill.\\} When suggestion spans and labels were correct, users with suggestions accepted them over 99\% of the time.  The median user without suggestions had a 89\% span recall and 86\% total recall on this set of annotations, confirming that users without suggestions do miss some examples. When spans were correct and labels were incorrect, the median user accepted suggestions 17\% of the time without modifications, and there were no observable difference between error rates between the modes (p=0.72, Kruskal-Wallis test). Namely, the presence of incorrect spans in \textit{Weakened} mode did not appear to induce additional mistrust in the incorrect labels; users in the \textit{Weakened} mode accepted incorrect labels at a similar 20\% median rate to users in \textit{Standard} and \textit{Augmented}. Further, the presence of the confidence indicators in the \textit{Augmented} mode did not make any noticeable impact on user's rate of modifying incorrectly suggested labels or their accuracy on the low-confidence subset. This matched their own feedback that "[the flags] didn't really affect the likelihood I accepted the suggestion" and that they do "not pay too much attention to the symbols."

Potentially due to the ease of accepting a pre-annotation in a single click, we observed a large difference between users in terms of how often they accepted incorrect label suggestions (a standard deviation of $0.14$). While the vast majority of users accepted between 10 and 20\% of incorrect suggestions, one user only accepted 8\%, and another accepted 58\%.  The user who accepted 8\% did not appear any better at the task; they had below-average total recall for their mode in Stage 1. Similarly, the user who accepted 58\% did not appear less competent at the task; they achieved above-average accuracy for their mode in Stage 1. This suggests that our domain experts reacted differently to automation, but this reaction is not directly a function of skill. From our user survey, we do note that users' perceptions of annotation accuracy do not necessarily reflect true underlying accuracy; for example, one user reported that they found the pre-annotated labels to be 95\% correct, despite their only accepting around 80\%. 

Though the majority of user errors were one-off, errors were not randomly distributed, and they indicated snap judgements based off concept name. For example, for a patient with a \texttt{persistent} cough, our system provided the incorrect label of \textit{persistence}, defined as \textit{mental perseverance}; this incorrect label was accepted by 10 users. For a patient suffering from \texttt{apical ballooning}, 6 users accepted the suggestion of \textit{balloon dilatation}. While by name alone, the concepts sound like plausible labels, in both cases they were of the incorrect concept category (e.g. \textit{balloon dilatation} is categorized as a \textit{Procedure} instead of a \textit{Problem}). While this category information is displayed to users, it seems they were not sufficiently engaged to utilize it. Our platform further provides a button for surfacing a label's definition, which was not taken advantage of by these users for either of these examples.

\subsubsection*{Users were slightly less accurate at getting rid of incorrect spans than incorrect labels.\\} On average, participants in the \textit{Weakened} mode kept 33\% of the 32 suggestions provided with incorrect spans, though the most conscientious user kept only 16\%. Trust in incorrect spans was strongly correlated with trust in incorrect labels ($\rho$=0.70, Spearman). However, like before, it did not hold any significant correlation with users' span accuracy in Stage 1  ($\rho$=-0.29), indicating that trust may be independent of user's competence at the annotation task. For example, the user who only accepted 16\% of incorrect spans had below average accuracy for their mode in Stage 1. Some of the provided incorrect spans included obviously incorrect selections such as \texttt{medical conditions. Exam} which spanned two sentences (kept by 2/5 users), or \texttt{of the superior segment branch}, which contained unnecessary prepositions (kept by 3/5 users). This indicates a decreased engagement. Other incorrect spans indicated concrete concepts, but were not clinical in nature, e.g. \texttt{sister}, and are therefore not supposed to be tagged. No users in any other mode chose to annotate any of extraneous spans mentioned here.

\begin{table*}[]
\begin{tabular}{|l|c|c|c|}
\hline
                & \textbf{\begin{tabular}[c]{@{}c@{}}Stage 1 Nontrivial\\ Span Recall\end{tabular}} & \textbf{\begin{tabular}[c]{@{}c@{}}Stage 2 Nontrivial\\ Span Recall\end{tabular}} & \textbf{\begin{tabular}[c]{@{}c@{}}User Survey on the Process of \\  Adding New Annotations\end{tabular}}                       \\ \hline
\textit{User A} & 63\%                                                                           & 37\%                                                                            & \begin{tabular}[c]{@{}c@{}}``I made sure to double check if there\\ were parts that were not annotated."\end{tabular}                                       \\ \hline
\textit{User B} & 71\%                                                                           & 56\%                                                                            & \begin{tabular}[c]{@{}c@{}}``I reviewed sections just in case I missed some... \\ but the marked sections were fairly comprehensive."\end{tabular}     \\ \hline
\textit{User C} & 68\%                                                                           & 58\%                                                                            & \begin{tabular}[c]{@{}c@{}}``Having pre-suggested parts actually made it easier to\\  scan the remaining unmarked parts for words to annotate"\end{tabular} \\ \hline
\textit{User D} & 63\%                                                                           & 62\%                                                                            & \begin{tabular}[c]{@{}c@{}}``[Pre-annotations] definitely freed up mental bandwidth \\ to allow me to spend more energy on the unmarked text."\end{tabular} \\ \hline
\end{tabular}
\caption{\label{tab:agency} Differences for four representative users between nontrivial span recall on Stage 1 (where they had no pre-annotations) and Stage 2 (where they did). Nontrivial span recall calculates the proportion of spans that users took the initiative to annotate that would not have been pre-annotated by our annotation suggestion algorithm. While the drop between Stage 1 and Stage 2 recall indicates that users took less initiative in practice, users believed they were more thorough in their exit surveys.}
\Description{The table includes four rows corresponding to Users A-D, and three columns containing nontrivial span recall from Stage 1, nontrivial span recall from Stage 2, and user comments describing their process of adding new annotations. In all four rows, recalls decrease, between 1\% and 26\%. However, users comment that they double check/review sections to ensure they're not missing anything, and that the presence of pre-annotations freed up their mental bandwidth to focus on unmarked sections.}
\end{table*}

\subsubsection*{Users with pre-populated suggestions exercised less agency in creating new annotations, but they noted the opposite in their exit surveys.\\} 
While users with pre-populated suggestions overall had slightly higher total recall than users without, they initiated the creation of fewer additional annotations. If we consider the subset of nontrivial annotations---annotations where suggestions were not provided and are therefore the most important to annotate---users with suggestions annotated fewer of these spans (median of 58\%) than users without suggestions (median of 74\%). Most striking was when we contrasted users to their own nontrivial recall from Stage 1; as a note, in Stage 1, there was no significant difference between nontrivial recall between the modes (p>0.4, two-sided Kruskal-Wallis). While Stage 1 did not contain suggestions, for a direct comparison, we restrict to the set of nontrivial annotations that \textit{would not have had} suggestions, had we applied the same suggestion algorithm. Users without suggestions increased on average 4\% between the two stages, indicating Stage 2 may have been slightly easier, but users with suggestions dropped on average 12\%. In an Aligned Rank transform test with factors of (i) stage and (ii) whether users received pre-annotations, there was a statistically significant effect of stage (p<0.005) and of the interaction between stage and mode (p=0.01).

Only a single user had a higher nontrivial span recall in the presence of suggestions. The difference was most pronounced in compound terms, where users without suggestions had a median of 64\% nontrivial span recall on compound terms, and users with suggestions only had a median of 31\%. For example, all users without suggestions tagged \texttt{RV thrombus} (compared to 46\% of users with suggestions) and \texttt{decompensated CHF} (compared to 50\% of users with suggestions).

Once again, this loss of agency was not correlated with users' prior accuracy and, qualitatively, users did not recognize it was happening. While a user's nontrivial recall results at Stage 1 and Stage 2 were highly correlated (Spearman's $\rho=0.76$, p<0.001), there was no correlation between their loss in agency (measured as difference in nontrivial recalls between stages) and their prior performance (Spearman's $\rho=0.25$. p>0.3). When prompted in their exit surveys, all stated that they believed the pre-annotations made it easier to tag the nontrivial spans. We display a few representative responses in Table \ref{tab:agency}, alongside their drop in nontrivial span recall. Even users (e.g. User C, User D) who were confident that they had greater bandwidth and performed better displayed a small drop in performance. 

\subsubsection*{Users were faster on annotations that came with pre-populated suggestions, but when labels were incorrect, they were slower than the users annotating from scratch. There was also a correlation between faster speeds and more errors. \\} 
Label time was calculated as the time between highlight and label choice (for annotations without suggestions) and time between a suggestion click and label choice (for annotations with suggestions).
76\% of all annotation spans in the gold standard were suggested to users as pre-annotations. Of these, 81\% had both a correct span and label. In the cases where both span and label were correct, users in suggestion modes only needed to verify the label was correct, and they were on average 30\% faster at label selection than users not in suggestion modes (a median of 5 vs 7 seconds). However, when labels are incorrect, users with suggestions were on average 5 seconds slower than users annotating from scratch; users also spent an average of 5 seconds on suggestions over incorrect spans. On examples without pre-annotations, users annotate at approximately the same speed as they had on comparative examples in Stage 1. We find that in our context the efficiency gains are slim in our context and depend greatly on their underlying accuracy. 

While users lauded the efficiency gains, we find that users with very high efficiency gains tended to be more inaccurate. In practice, some speedy users would quickly accept pre-annotations in less than a second; video revealed that their primary tactic was to first quickly deal with pre-annotations, and then go back and scan through the remaining text. In contrast, the user with the second lowest rate of accepting erroneous label suggestions was also one of the slowest, taking almost twice as long as the median. In exit surveys, users believed the pre-annotations were great for efficiency, noting that they were ``much faster," ``more efficient," and ``cut down time significantly". Empirically, however, their utility is less clear-cut. 

\section{Discussion}
The results of our study suggest implications for the appropriate amount of automation to include in our and similar platforms. We found that our domain experts were relatively adept at dealing with incorrect labels: both searching for additional labels when recommendations failed, and modifying incorrect labels in pre-populated annotations. This would indicate that there is minimal downside to the inclusion of label recommendations in the platform UI, regardless of whether it is presented in a list format, or as a single prediction. In contrast, in Stage 2, users were slightly less likely to get rid of incorrect spans in pre-populated annotations, even when spans were clearly incorrect. This would indicate that the set of pre-populated annotations would need to prioritize high precision (surfaced annotations correspond to true spans) over high recall (most gold standard annotations are surfaced). Additionally, we found the presence of pre-populated annotations redirected users' attention and caused a loss of agency, and such techniques should be employed with caution. 

Further, our study revealed some characteristics of our domain experts. First, they did not realize they had ceded agency, and this underscores that they are not fully cognizant of their interactions with automation. Therefore, their own qualitative conclusions on how they operate in teams are generally insufficient; this aligns with recent work from Buçinca and Lin et al.~\cite{Bucinca2020}. Second, we found that the negative impacts of automation were user-specific and correlated across aid modalities. However, these impacts were independent of a user's competency at the task. 

The optimal decision regarding how to integrate decision support into a system may depend on the downstream use case. Counter to prior concerns, we found that decision support increased users' average recall. Therefore, if the output of a task is being directly used for decision making, then the full inclusion of decision support may be useful. However, if the output of a task is to be used for further model training, there would be a possible fear of creating a feedback loop with pre-populated annotations, since we found that users are less likely to annotate concepts that the machine missed. When fed back into models, users' output may erroneously confirm the machine's decision. 

The clinical text annotation task involves a rich set of subtasks and decisions that allowed us to probe questions of trust and agency in expert decision makers. However, our findings do not necessarily extend to simpler settings. For example, due to our large potential label space, it was implicitly clear to users that the model is presenting only a subset of possibilities. In contrast, if a domain expert only has to make a decision with a binary outcome (e.g. \textit{Does a pathology image indicate cancer?}), the user may develop a different dynamic with decision support, since the problem is more constrained. That being said, binary problems like image classification often involve smaller implicit subtasks (e.g. \textit{Does this patch of the image display cancer?}). As automated methods attempt to focus users' attention on the subtasks they deem most relevant, our findings on agency may still extend. 

Other limitations include the low-stakes setting of our study task. Since we did not deploy in a live setting, and there was no consequence for incorrect actions, users may not have been as careful as they would have been under greater pressure. Additionally, we found that presenting model confidence had little-to-no-impact, and users admitted to not using the confidence flags. Other UI techniques might have made these confidences more salient to users. 

Several techniques could mitigate the shortcomings of human-AI teams from a system design perspective, both for our task and more broadly. For example, we could withold pre-annotated suggestions during the initial round of annotation; after, a second annotator could incorporate any missing spans from pre-annotated suggestions. This may dampen the effect of the observed loss of agency. We could also experiment with pre-populated span suggestions without pre-selected labels, which could force users to remain critically engaged. Another option would be to continually engage the user with regular feedback; currently, they only receive direct feedback during the tutorial phase. Possibilities include (i) purposefully planting a fraction of erroneous recommendations and suggestions and alerting the user to when they've accepted one and (ii) alerting the user when they skip over a span they should have annotated. Additionally, the amount of training may affect outcomes, and we could investigate whether more rounds of early tutorial feedback would lead users to better mentally characterize the shortcomings of the decision support. 

Further, we found that automation had widely varying impacts on users and ceded agency tended to correlate across modalities. One simple option would be to filter users with high misplaced trust in automation or to predict when user log data indicate they may be running on autopilot. However, that is not always a realistic solution, and our work further found that susceptibility was not correlated with competency or skill. Therefore, another future direction would be to understand users' susceptibilities by their early results (e.g., in the tutorial) and then adjust the level of automated decision aid provided accordingly. 

Instead of just attempting to fix the automation-induced noise at the point-of-annotation, another path would be to design machine learning algorithms that anticipate and adapt to the noisy data. Natarajan et al. showed that in the presence of noise, binary machine learning classifiers can still be successfully trained, if the patterns of noise are well-characterized and below random (e.g. there is an estimate of what fraction of the time a specific outcome may be incorrect) \cite{Natarajan2013}. In these cases, models can be trained by re-weighting their objective function by a factor dependent on the probability of user error. In our case, we would need to adapt such algorithms to account for probabilities of users accepting incorrect spans or skipping certain categories of annotations, for example. Given our observed empirical results, we posit that machine learning methods designed to overcome the pitfalls of human-AI teams are an important area of future study \cite{ratner2017snorkel}.

\section{Conclusions}
In this work, we presented a self-contained platform in which users can annotate spans of text and map them to large label spaces. We used this platform to study the impact of decision aid on domain experts via empirical lab studies on clinicians (n=18) over extended periods of use. More broadly, our platform enables efficient annotation of text documents and could help scale data set creation in a domain where annotated data set sizes have been historically small.

On the whole, we found that our domain experts remained appropriately skeptical of label recommendations, and they formed an intuition for when further searching was required. Similarly, they mostly recognized when pre-populated labels were incorrect. As a result, the introduction of automatic label recommendations is unlikely to lead to significant bias. Unfortunately, our domain experts do fall susceptible to handing over agency to algorithms. Without them realizing it, the presence of pre-populated suggestions leads them to lose critical engagement in the task and add fewer new annotations than each had previously. Given that these new annotations are the ones that provide us the most new signal for training models, we would be in danger of models being hampered in their training process. 

As automation becomes incorporated into more and more decision processes, it becomes paramount for us to understand how automation affects expert decision makers. As we found with varied susceptibility among users, issues of algorithmic trust and agency extend far past user confidence and expertise to a more intrinsic behavior. As we do in this work, understanding,  characterizing, and quantifying that behavior in complex, real-world tasks is an important first step. It informs the design of both user interfaces and machine learning systems that can optimally combine the strengths of humans and AI and mitigate their joint shortcomings.

\begin{acks}
We would like to thank Chloe O'Connell, Yasmin Fatemi, and Zeshan Hussain for providing iterative feedback on our annotation platform; Divya Gopinath, Christina Ji, Hunter Lang, and Hussein Mozannar for editorial help; Nathalie Vladis for her statistics aid; and the reviewers whose suggestions strengthened our paper. We would also like to deeply thank all the clinicians who participated in the study despite their busy schedules.
\end{acks}

\bibliographystyle{ACM-Reference-Format}
\bibliography{acmart}

\appendix

\end{document}